\documentclass[12pt]{iopart}

\usepackage{graphicx}
\date{\today}

\begin{document}

\title{Triple differntial cross section for electron impact ionization of hydrogen atom}

\author{Hari P. Saha}

\address{Physics Department, University of Central Florida, Orlando, FL 32816. }
\ead{Haripada.Saha@ucf.edu}
\begin{abstract}

 The electron impact ionization of atomic hydrogen is calculated for incident elrctron energy 76.46 eV.  The Hartree-Fock approximation is used to calculate the initial state which includes both bound and continum wave functions.  The final state continuum electron wave functions are obtained in the potential of hydrogen ion.   The interaction between the two final state continuum electrons is approximated with the screening potential determined variationally.
\end{abstract}

\maketitle

\section{\label{sec:intro}Introduction}

 It is of fundamental importance to study triple differential cross sections for the
 ionization of atoms both theoretically and experimentally, as TDCS provide valuable information
 about the structure  of atoms between the three bodies.  Several studies have been performed earlier
 on several targets using different approximations.  Hydrogen has been chosen for detailed investigation because of their simple structure.  Among the various theoretical approaches, the convergent close coupling method (CCC), the time dependent close coupling (TDCC) method and the exterior complex scaling (ECS) approach are considered to produce accurate results.  The multichannel R-matrix theory, the distorted wave method and the distorted partial wave approach also produce reliable results. The TDCS at low and intermediate energies are studied by Braunner et al. using the final state wave functions which is correct asymptotically.

 Previously we investigated TDCS of H atom by electron impact at several incident energies for the simplified case when the two final state continuum electrons leave in the opposite direction using our extended MCHF method.  The initial
 state is calculated in the HF approximation and the final state wave function are obtained in the HF and the variationally determined screening approximation for both equal and unequal sharing of excess energy by the two final state continuum electrons.  Both HF and the SP results are found to be very close to each other indicating that when the two final state continuum electrons leave in the opposite direction.  The results are found to be very good with relative experimental measurement and the other accurate calculations.

 Non-perturbative methods e.g ECS, CCC, and TDCC approaches are known to provide more accurate results only for ionization of simple targets such as H and He than those provided by the SP approximation.  Although nonpertubative  methods can explain the experimental measurement very accurately but they have limitations for multi-electron targets  or complex atoms. The application of the present, one of the perturbative, on the otherhand, can be applied to multi-
electron targets, when non-perturbative methods have difficulty.\

The application of the present approach, one of the perturbative approaches for electron atom ionization privide a first step in testing ionization from complex targets for which non-perturbatve methods have not been applied.  So far, our calculations with the SP approximation are performed for a simple geometry when the  two final state continuum electrons are leaving in the opposite direction.  All calculations found that results obtained in the SP approximation provide very good results with the experiment and the other accurate calculation.  As already mentioned, investigation on TDCS for electron impact ionization of H and He atoms using the HF and the SP approximation for $ \theta_{12}= \pi$  symmetry have been reported.  It has been found that the results are very encouraging.  It was also found that in the case of H and He the electron correlation between the two final  state continum elecons are small when the two continuum electrons leave in the opposite direction.  In this paper we investigated the application of the SP approximation to study the TDCS of H for incident electron energy 76.46 eV with two final state electrons sharing 31.43 eV each for other kinematics and collision geometry which will prove the collision dynamics.

The study of electron impact ionization of atoms has been the
subject of fundamental interest and importance both theoretically
and experimentally as they provide valuable information about the
structure of atoms and electron dynamics between the three bodies.
 Several studies have already been made earlier on several targets
 using different approximations.  Among many targets hydrogen and
 helium have been investigated in detail because of their simple
 nature.  Theoretically few approximations which produce accurate
 results are the convergent close coupling (CCC) method \cite{IB}, the
 time-dependent close coupling method \cite{CPR} and the exterior complex
 scaling \cite{RBI} approach.  Few other theoretical methods which
 provide reliable results are the multichannel R-matrix theory \cite{BB},
 distorted wave method \cite{MCS} and the distorted partial wave approach \cite{PS}.  In
 addition , Braunner et al.\cite{BBK1,BBK2} used the exact asymptotically correct
  final state wave function to study triple differential cross
  section(TDCS) for electron impact ionization of H and He at low
  and intermediate energies.  Temkin \cite{ATEM} studied theoretically the
  behavior of electron impact ionization of atoms
  by developing the Coulomb dipole theory.  All these theoretical
  models paid their attention to improve the final state wave
  function more accurately.  We reported \cite{SAHA1,SAHA2} earlier the results of
  our study for the TDCS of H by electron impact at several
  incident energies for the case when the two final state
  continuum electrons leave in the opposite direction using the
  Hartree-Fock (HF) and the Screening potential (SP) approximations.
  Recently we performed calculations
   on low energy electron impact ionization of helium atom using
   MCHF method \cite{SAHA4}.  In another recent paper \cite{SAHA3}, we considered target
   correlation and polarization effects on electron impact ionization
   of helium atoms.  We found that polarization of the target by the
 incoming electron has considerable effects on the electron impact
 ionization of He atom.   Experimentally , Ren et al. \cite{REN} has made both
 experimental and theoretical study on electron impact ionization
 of He by the 70.6 eV incident electron, which covers entire solid
 angle for the emitted electron and the collision kinematics.  They
 compared the absolutely normalized triple differential experimental
 cross section with the  theoretical calculation obtained by CCC
 and TDCC methods and found excellent agreement.

In this paper we are interested in the calculation of the TDCS of
H at 76.46 eV incident energy for equal  sharing of 62.86
eV  excess energy.  It is very important to calculate the initial
and the final state wave functions accurately to obtain accurate
TDCS. As mentioned earlier, most of the methods were designed
mainly to treat the final state correlation accurately.  In this
work we have carried out calculation of the TDCS of H-target at
76.46 eV incident electron energy for the case when the two final
state outgoing electrons share the 62.46 eV excess energy
equally.  The main purpose of this investigation is
to test the screening potential approximation on TDCS at the
incident energy and to compare our results with the available experimental
and the other theoretical data. The final state interaction
potential between the two final state continuum electrons is
included using the variationally determined screening potential
(SP).

\section{\label{sec:Theo}Theory}

 A.  Triple Differential Cross Sections \\

The description of the MCHF theory of electron impact ionization
of atoms is provided in earlier papers \cite{SAHA1,SAHA2}. Briefly
, the triple differential cross section for electron impact
ionization of atoms is given by \cite{PS}

\begin{eqnarray}
{\frac{d^3\sigma}{dE_2d\Omega_1d\Omega_2}} = \frac{(2\pi)^4}{k}k_1
k_2 {|< {\Psi_f}^-| {V}| {\Phi_i}^+ >|}^2
\end{eqnarray}

\noindent where $\vec{k}$ is the momentum of the incident electron
and $\vec{k_1}$ and $\vec{k_2}$ are the momenta of the two
continuum electrons in the final state.  ${\Phi_i}^+$ and
${\Psi_f}^+ $ represent the initial and final state wave functions
of the system  respectively. $E_i = \frac{{k_i}^2}{2}$ is the
kinetic energy of the $i^{th}$ final state continuum electron.
The solid angles $d\Omega_1$ and $d\Omega_2$ are associated with
the two final state continuum electrons. The perturbation V is the
difference between the exact Hamiltonian and the approximate
Hamiltonian used to construct and describe approximately the
initial state ${\Phi_i}^+ $ and is approximately defined as
\cite{PS}

\begin{eqnarray}
V = \sum_{i=1}^{N} \frac{1}{| {r_{N+1}-r_i}|}- { V_{HF}}^{N+1}
(r_{N+1})
\end{eqnarray}

\noindent where the first term on the right hand side of this
equation is the coulomb interaction between the incident electron
and the N-target electrons and the second term is a
multiconfiguration Hartree-Fock approximation to this interaction
which is used to construct the initial state $ {\Phi_i}^+$.  The
initial state ${\Phi_i}^+ $ is described by the orbital and spin
angular momentum $L_0$ and $S_0$ of the target and by the momenta
$\vec{k}$ and orbital angular momentum $l$ of the incident
electron.  The final state wave function $ {\Psi_f}^- $ is
characterized by the orbital and spin angular momenta $L_c$ and
$S_c$ of the (N-1) electron of the core ion and by the momenta
$\vec{k_1}$ and $\vec{k_2}$ and by orbital angular momenta
$l_1,l_2$ of the two continuum electrons. \\

 Using the partial wave expansion of the incident electron and each
of the two final state continuum electron wave functions we expand
the initial state ${\Phi_i}^+ $ and the final state ${\Psi_f}^- $
wave function for the (N+1) electron system.  The triple
differential cross section then reduces to

\begin{eqnarray}
    {\sigma_{He}}^{(3)}= \frac{4\pi}{k^2[L_0][S_0]}\sum_{S}
    \mid   \sum_L (2L+1)A (LS \hat{k_1}\hat{k_2})\mid ^2
\end{eqnarray}

\noindent where
\begin{eqnarray}
A(LS \hat{k_1}\hat{k_2}) = \sum_{l_1l_2}\sum_{m_1m_2}i^{l+l_1+l_2}
   e^{ i(\delta_l+\sigma_{l_1}+\delta_{l_1}+\sigma_{l_2}+\delta_{l_2})
     } \nonumber \\
\left(\begin{array}{ccc}l_1&l_2&L\\m_1&m_2&0\end{array}\right)
Y_{l_1m_1}(\theta_1,\phi_1) Y_{l_2m_2}(\theta_2,\phi_2)<\psi_f
\vert V \vert \psi_i>
\end{eqnarray}

\noindent with [x] = (2x+1)  $$\psi_i \equiv
\Psi_i((L_0l)L_TM_T(S_0\frac{1}{2})S_TM_{S_T}),$$ $$\psi_f \equiv
\Psi_f([L_c(l_1l_2)L]L_TM_T[S_c(\frac {1}{2}\frac
{1}{2})S]S_TM_{S_T}) $$

 \noindent Here L and S are the
orbital and spin angular momenta of the final-state continuum pair
and $L_TM_T $ and $S_TM_{S_T} $ are the total orbital and spin
angular momenta of the system.  \\

B. Wave functions for the Continuum electrons  \\

The multi-channel multi-configuration Hartree-Fock (MCHF) method
is described earlier \cite{SM}.  The total wave function in the HF
approximation [11] at energy $E=E_i + {k^2/2}$ and term value LS
can be expressed as \cite{SM}

\begin{eqnarray}
\Psi_E &=& \Phi(\gamma_iL_iS_i;N)F_{k_il_i}
%\nonumber\\
 %  &+&
%\sum_{j=1}^{N_j}C_j\Phi_j(\gamma_jL_jS_j;N+1)
\end{eqnarray}
\noindent
 where $ \Phi(\gamma_iL_iS_i;N) $ represents N-electron target wave function
having energy $E_i$, configuration $\gamma_i$ and the term $L_i$
and $S_i$ coupled with a single electron wave function
$F_{k_il_i}$ having energy $\frac{1}{2}{k_i}^2$ (in atomic units)
and orbital angular momentum $l_i$ to form an antisymmetric
configuration for the (N+1) electron system with a designated term
value.  The above wave function is defined in terms of a set of
radial functions $P_i(r), i= 1,......,m.$  As for example,
$F_{k_il_i}=\frac{P_i(r)}{r}Y_{l_im_i}(\theta_i,\phi_i)\chi_{m_s}$
where $Y_{l_im_i}(\theta_i,\phi_i)$ is the spherical harmonic and
$\chi_{m_s}$ is the spin function.  The set of radial functions
$P_i(r), i=1,....,m_t $ describing the targets are obtained from
the HF bound state calculations for the targets and are kept
fixed.  The set of radial functions describing the continuum
orbitals are determined variationally. These radial functions are
the solutions of the integro-differential equations of the form
\cite{SM},

\begin{eqnarray}
 [\frac{d^2}{dr^2}+\frac{2Z}{r}-\frac{l_i(l_i+1)}{r^2}]P_i(r) =
 \frac{2}{r}[Y_i(r)P_i(r)+  \nonumber\\
 X_i(r)+I_i(r)]+\sum_{i'}\epsilon_{ii'} P_{i'}(r)
\end{eqnarray}

\noindent which has the same form as the Hartree-Fock equation for
a singly occupied orbital of a bound state system, the only
difference being the specified binding energy, $ \epsilon_{ii} =
\frac{k^2}{2}$ and the boundary condition at infinity.

\noindent In this equation $\frac{2}{r}Y_i(r)$ is a part of direct
potential, $\frac{2}{r}X_i(r)$ is the exchange function and
$\frac{2}{r}I_i(r)$ represents terms arising from interactions
between the configurations.  The off-diagonal energy parameter
$\epsilon_{ii'}$ are related to Lagrange multipliers that ensure
orthogonality between the continuum and the bound electrons of the
target having the same symmetry.  These operators have their usual
meanings as for bound state problems.

In the single channel case, the radial function $P_i(r)$ satisfies
the boundary conditions,

\def\buildrel#1\below#2{{\mathrel{\mathop{\kern0pt
#2}\limits_{#1}}}}
\begin{eqnarray}
P_{i}(r) ~{\buildrel r\to 0 \below \rightarrow}~ r^{l+1}~,
  ~~~
P_{i}(r) ~{\buildrel r\to +\infty \below \rightarrow}~
\sqrt{\frac{2}{\pi k_i}}\sin(k_ir-\frac{l_i\pi}{2}+\delta_{l})
\end{eqnarray}
if the target is an atom and
\begin{eqnarray}
P_{i}(r) ~{\buildrel r\to +\infty \below \rightarrow}~
\sqrt{\frac{2}{\pi
k_i}}\sin(k_ir-\frac{l_i\pi}{2}+\frac{q}{k_i}\ln{2k_ir}+\sigma_{l}+\delta_{l})
\end{eqnarray}
if the target is an ion. Here $\sigma_l
=\arg[\Gamma(l_i+1-\frac{iq}{k_i})]$ is the
coulomb phase shift.  q = Z-N is the net charge of the ion. \\

The integro-differential equation (6) is solved numerically by the
iterative method similar to the bound state problem. The
self-consistent field procedure is applied to compute the
continuum wave functions.  The continuum radial function is
normalized by fitting the computational values at two adjacent
points to the regular and irregular Bessel or Coulomb functions
depending on the target as soon as the region is reached where the
direct and exchange potentials are vanishingly small.  This may be
at a considerably smaller value of r than the asymptotic form
represented by the boundary conditions specified in equations (7)
and (8.).   \\

3.  Approximations used to calculate initial and the final
state wave functions.  \\

In this paper, we have considered the HF approximation to
calculate the incident electron wave function.  For the final
state continuum electron wave functions the HF and the screening
potential (SP) approximations are used.           \\

\noindent Initial State:  In the present case, the target
considered is the hydrogen  atom.  We first calculated the initial
state hydrogen atom wave function in the Hartree-Fock (HF)
approximation.  As already mentioned, the initial state wave
function ${\Phi_i}^+$ is expanded in terms of antisymmetrized LS
coupled wave function of the N electron target and the single
electron wave function of the incident electron.  The continuum
radial wave functions are calculated by solving the
integro-differential eqn(6) with the HF potential of the target
hydrogen atom under the specified boundary conditions, where the
target electron wave function is kept frozen at incident electron
energy 76.46 eV for angular momentum from $ l=0 $ to $l=6$.  \\

\noindent Final State: To examine the effect of electron
correlation in the final state, the final state continuum electron
wave function is calculated in two approximations.  (i) the HF
approximation and (ii) the screening potential (SP) approximation.
It should be mentioned that the screening potential approximation
is an approximation to the actual Coulomb interaction potential
between the two continuum electrons.  It accounts for partial
electron correlation. In the screening potential approximation the
exact Coulomb interaction between the two continuum electrons in
the final state is replaced by a variationally determined angle
dependent screening potential due to mutual screening
\cite{RS,JZF} of the nucleus by the ejected electrons using
effective charges which satisfy proper asymptotic boundary
conditions.  On the other hand, in the HF approximation we ignored
this interaction between the two final state continuum electrons.
The difference between the wave functions calculated with these
two approximations will determine the effect of distortion in the
final state wave functions.  The screening potential for the two
continuum electrons are determined by the effective screening
charges $\Delta_1 $ and $\Delta_2 $ which are obtained by the
condition \cite{PS,RS,JZF}

\begin{eqnarray}
\frac{Z_T - \Delta_1}{k_1} + \frac{Z_T - \Delta_2}{k_2} =
\frac{Z_T}{k_1} + \frac{Z_T}{k_2} - \frac{1}{\vec{k}_1-\vec{k}_2}
\end{eqnarray}

\noindent where $ Z_T $ is the net asymptotic charge of the
ionized target.  The effective screening charges which satisfy the
above relation are obtained as \cite{PS,RS,JZF},

\begin{eqnarray}
      \Delta_i = \frac{(\vec{k}_i.\vec{k}_{ij})k_i}{{k_{ij}}^3}   ~~~  (i=1,2).
\end{eqnarray}

\noindent where $\vec{k}_{ij}=\vec{k}_i - \vec{k}_j,  j \neq i, $
esium atomesium atomwith $k_{ij} = | \vec{k}_{ij} | $ \\

The wave function for each of the final state continuum electrons
in the SP approximation
 are calculated using the same numerical procedure as adopted in the
 multi-configuration Hartree-Fock method
\cite{SAHA3,SAHA4} for bound
 and continuum electrons at each relative angle between the two
 continuum
  electrons ejected at  equal energy  for the
 angular momentum $l = 0$ to $ l=6 $ for the  partial wave $L =
 1-6$ and
$S = 0 - 1$.

\section{\label{sec:Resu}Results}

\begin{figure}[h]
\vspace{1.cm}
\includegraphics[width=8cm]{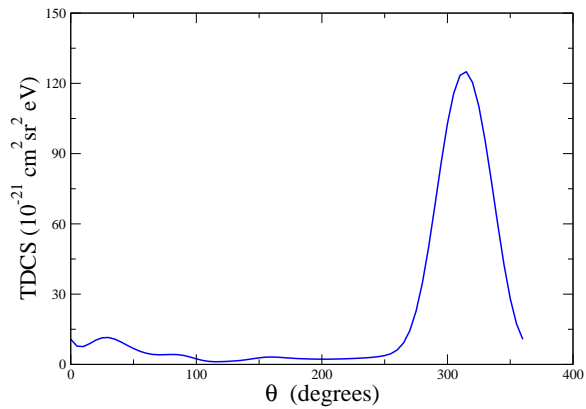}
\vspace{0.3cm}\caption{Comparison of present triple differential
cross sections of He in the scattering plane for equal energy
sharing $E_1=E_2 = 31.43 eV$, as a function of ejected electron angle
$(\theta_1)$ for the scattered electron angle $\theta_1$ fixed at
$\theta_1  = 30^0$.}
\end{figure}

In this case, we discuss the triple differential cross section for
electron impact ionization of
hydrogen atom with the initial state
calculated in the HF approximation at 76.46 eV incident electron
energy and the final state in the HF and the screening potential
(SP) approximation with the excess energy 62.86 eV shared equally
 by the two final state continuum electrons at a
 scattered electron angle $\theta_1 = 30^0$ in both inplane and out of
plane.  \\

  Equal energy:  $E_1= E_2= 31.43 eV $    \\

 Figure represents the TDCS calculated in the HF and the SP
 approximations for the 62.86 eV excess energy sharing equally
 between the two final state electrons in the scattering plane
  at fixed scattered electron angle $\theta_1 = 30^0$.

\section{\label{sec:con}Conclusion}

We studied the ionization of hydrogen atom by 76.46 eV electron in the HF
and the SP approximation .  We calculated the triple differential
cross sections
with the final state continuum electrons sharing 62.86 eV excess
energy equally  for a fixed scattered electron
angle.    We use HF approximation for the
initial state and both HF and SP approximations for the final
state.  The interaction potential between the two final state
continuum electrons is approximated by variationally determined
screening potential.  It only accounts for partial potential and
can not be considered as a correct electron correlation between
two final state continuum electrons.  We could not compare the present
results with experiment as the available experimental results are not reliable.

%\begin{acknowledgments}
\section*{Acknowledgements}
We are very grateful to  Dr. J. Colgan for sending us their
calculated data in electronic form.  We express our thanks to Drs.
Klaus Bartschat, Igor Bray, T.N. Resigno , M.S. Pindzola, F.J.
Robicheaux and T.W Gorczyca for their help.

%\end{acknowledgments}

\end{document}